\documentclass[3p,times]{elsarticle}

\usepackage{ecrc}
\usepackage{amsmath}
\usepackage{slashed}
\usepackage{hyperref}

\volume{00}

\firstpage{1}

\journalname{Nuclear Physics A}

\runauth{}

\jid{npa}

\jnltitlelogo{Nuclear Physics A}

\usepackage{amssymb}

\biboptions{square,comma,numbers,sort&compress}

\usepackage[figuresright]{rotating}

\def\sss{\scriptscriptstyle}

\newcommand{\Tr}[1]{\textrm{Tr}\left[#1\right]}

\def \cf {C_{\sss F}}
\def \nc {N_c}

\def \nf {N_f}

\def \dr {d_{\sss R}}
\def \da {d_{\sss A}}

\def \bk {\mathbf{k}}

\newcommand{\Tint}[1]{{\hbox{$\sum$}\!\!\!\!\!\!\!\int\,}_{\!\!\!\!\raise-0.9ex\hbox{$\scriptstyle{#1}$}}}

\def \cc{{\mathcal C}}

\def\siml{{\ \lower-1.2pt\vbox{\hbox{\rlap{$<$}\lower6pt\vbox{\hbox{$\sim$}}}}\ }}
\def\simg{{\ \lower-1.2pt\vbox{\hbox{\rlap{$>$}\lower6pt\vbox{\hbox{$\sim$}}}}\ }}

\def \bk {\mathbf{k}}

\def \als {\alpha_{\mathrm{s}}}

\def \m2   {\mu^{2 \epsilon}}

\def\siml{{\ \lower-1.2pt\vbox{\hbox{\rlap{$<$}\lower6pt\vbox{\hbox{$\sim$}}}}\ }}
\def\simg{{\ \lower-1.2pt\vbox{\hbox{\rlap{$>$}\lower6pt\vbox{\hbox{$\sim$}}}}\ }}

\def\pp {p_\perp}
\def\xp {x_\perp}

\def\qp {q_\perp}

\def\bqp {\mathbf{q}_\perp}

\def\mm {m_\infty^2}
\def\md {m_{\sss D}}
\def\dgk{\frac{d\Gamma_\gamma}{d^3k}\bigg\vert}

\def\ddgkv{\frac{d \delta \Gamma_\gamma}{d^3k}\bigg\vert}
\def\2to2{{2\leftrightarrow 2}}
\def\soft{{\mathrm{soft}}}
\def\hard{{\mathrm{hard}}}
\def\coll{{\mathrm{coll}}}
\def\sc{{\mathrm{semi-coll}}}
\def\LO{{\mathrm{LO}}}
\def\NLO{{\mathrm{NLO}}}

\def\k{{\bf{k}}}
\def\f{{\bf{f}}}

\def\x{{\bf{x}}}
\def\OO{{\mathcal{O}}}

\def\nfd{n_{\!\sss F}}

\begin{document}

\begin{frontmatter}



\dochead{}

\title{Next-to-leading order thermal photon production in a weakly-coupled plasma}


\author{Jacopo Ghiglieri}

\address{McGill University, Department of Physics,\\3600 rue University, Montreal QC H3A 2T8, Canada}

\begin{abstract}
We summarize the recent determination to next-to-leading order of the thermal photon rate at weak coupling. 
We emphasize how it  can be expressed in
terms of gauge-invariant condensates on the light cone, which are amenable to
novel sum rules and Euclidean techniques.  For the phenomenologically interesting
value of $\als=0.3$, the NLO correction represents a 20\% increase and has a
functional form similar to the LO result.
\end{abstract}


\end{frontmatter}


\section{Introduction}
\label{sec_intro}
Photons have long been considered a key hard probe of the medium produced in heavy-ion
collisions. Experimentally, there are now detailed data on real photon production at
RHIC \cite{Adare:2008ab,Adare:2011zr,Afanasiev:2012dg} and 
the LHC \cite{Lee:2012cd,delaCruz:2012ru,Milov:2012pd,Steinberg:2012tv,Wilde:2012wc}. Photons arising
from meson decays following hadronization are subtracted from the data experimentally; 
theoretically one then needs to deal with several sources:
\emph{``prompt'' photons}, produced in the scattering
of partons in the colliding nuclei, 
\emph{jet photons}, arising from the interactions and fragmentations of jets,
\emph{thermal photons}, produced by interactions of the (nearly) thermal constituents of 
the plasma and 
\emph{hadron gas photons}, produced in later stages.

In this contribution we will concentrate on thermal photons. Speficically, we will
consider a weakly-coupled, infinitely extended, static and equilibrated medium and, 
after briefly summarizing the leading-order  perturbative result 
\cite{Arnold:2001ba,Arnold:2001ms}\footnote{Partial LO results \cite{Shen:2013cca} for
non-equilibrated, anisotropic media have been presented at this conference 
in \cite{Chuntalk,Ulitalk}. Results for dileptons, in and out of equilibrium, have been presented in 
\cite{Laine:2013vma,Mikkotalk} and \cite{Gojkotalk} respectively.}, we will illustrate the NLO, i.e. relative $\OO(g)$,
 calculation
presented in full detail in \cite{Ghiglieri:2013gia}. Its motivations are both phenomenological, i.e. an improved
 knowledge of the rate and, importantly, a first estimate of its associated theory uncertainty, and theoretical.
Perturbation theory at finite temperature is known to suffer from slow convergence, at least for thermodynamical
quantities; with the present calculation we aimed at exploring the largely uncharted territory of dynamical, time-dependent
quantities beyond leading order. A posteriori, the extensive factorizations we found and the possibility of greatly
simplifying the calculation through Euclidean and sum rule technologies make  this calculation a pattern for future ones
of related dynamical quantities, such as jet energy loss and transport coefficients.

\section{Overview of the calculation}
\label{sec_lo}
The photon production rate is given at leading order in $\alpha$ by
\begin{equation}
	\label{defrate}
   (2\pi)^3\frac{d\Gamma_\gamma}{d^3k}
        =\frac{1}{2k}\,\int d^4X e^{-i K\cdot X}\left\langle J^\mu(0)J_\mu(X)\right\rangle,
\end{equation}
which relates it to the backward Wightman correlator of the
electromagnetic current $J^\mu=\sum_{q=uds}e_q\bar q\gamma^\mu q$. $K=(k,\bk)=(k,0,0,k)$ is the lightlike momentum 
of the photon,\footnote{
Here and throughout this contribution capital letters stand for four-vectors, lowercase 
italic letters for the modulus of the spatial three-vectors and the metric signature 
is $({-}{+}{+}{+})$, so that $P^2=p^2-p^2_0$.} which we choose to be oriented along the $z$ axis. We furthermore assume 
$k\simg T$, which is the validity region of the LO and NLO calculations.
We will work perturbatively in the strong coupling $g$, meaning that we
treat the scale $gT$ (the soft scale) as parametrically smaller than the
scale $T$ (the hard scale).

\begin{figure}[ht]
\begin{center}
\includegraphics[width=8cm]{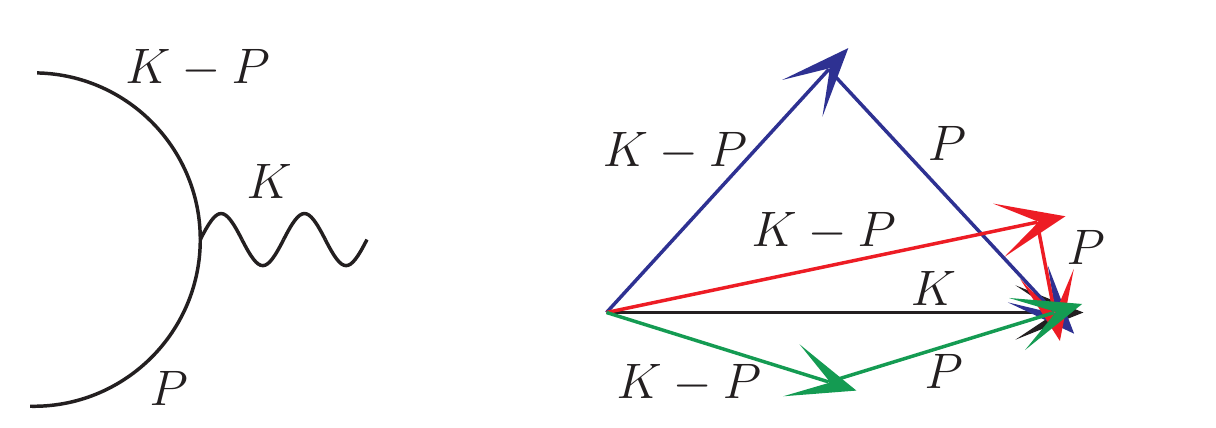}	
\end{center}
\vspace{-0.5cm}
\caption{Left: momentum assignments at one of the two current insertions. Right: the hard (blue), soft (red) and collinear (green) scalings are schematically represented.}
\label{fig_moms}
\end{figure}

\begin{figure}
\begin{center}
\includegraphics[width=8cm]{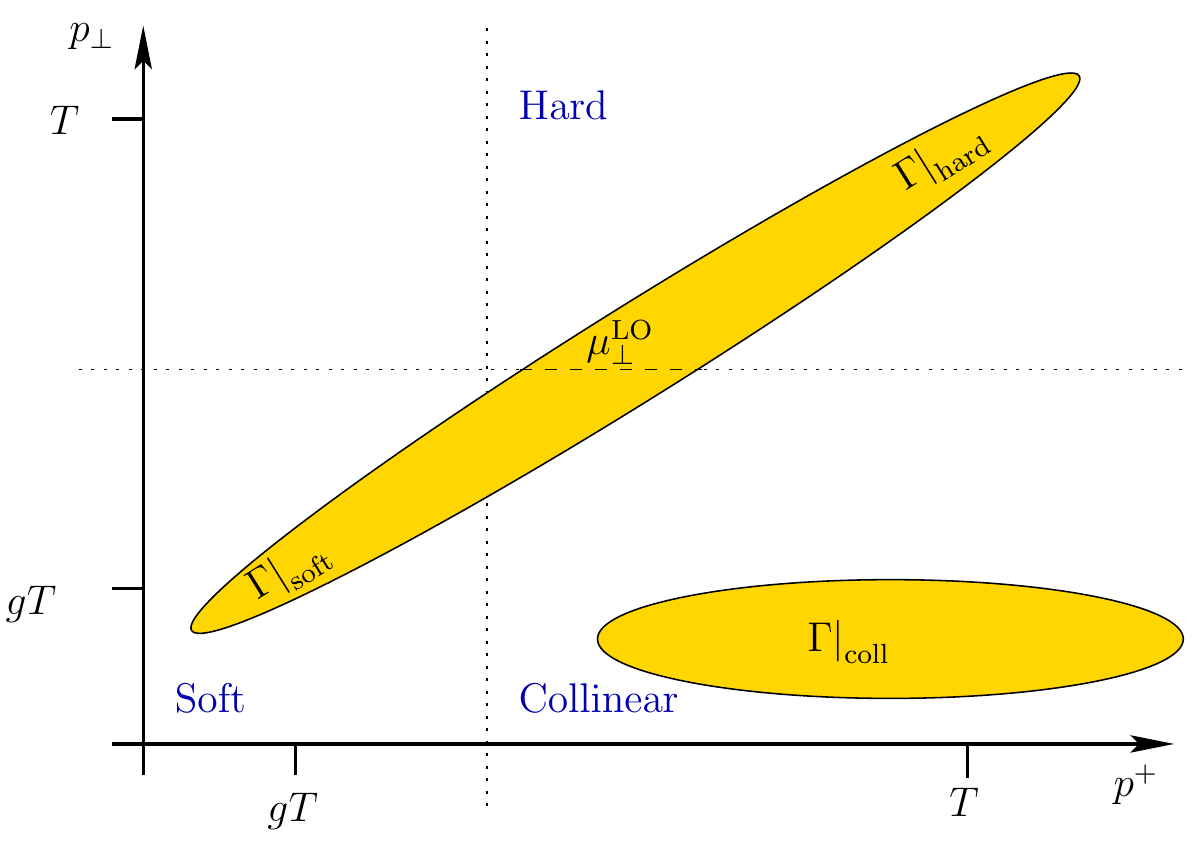}	
\includegraphics[width=8cm]{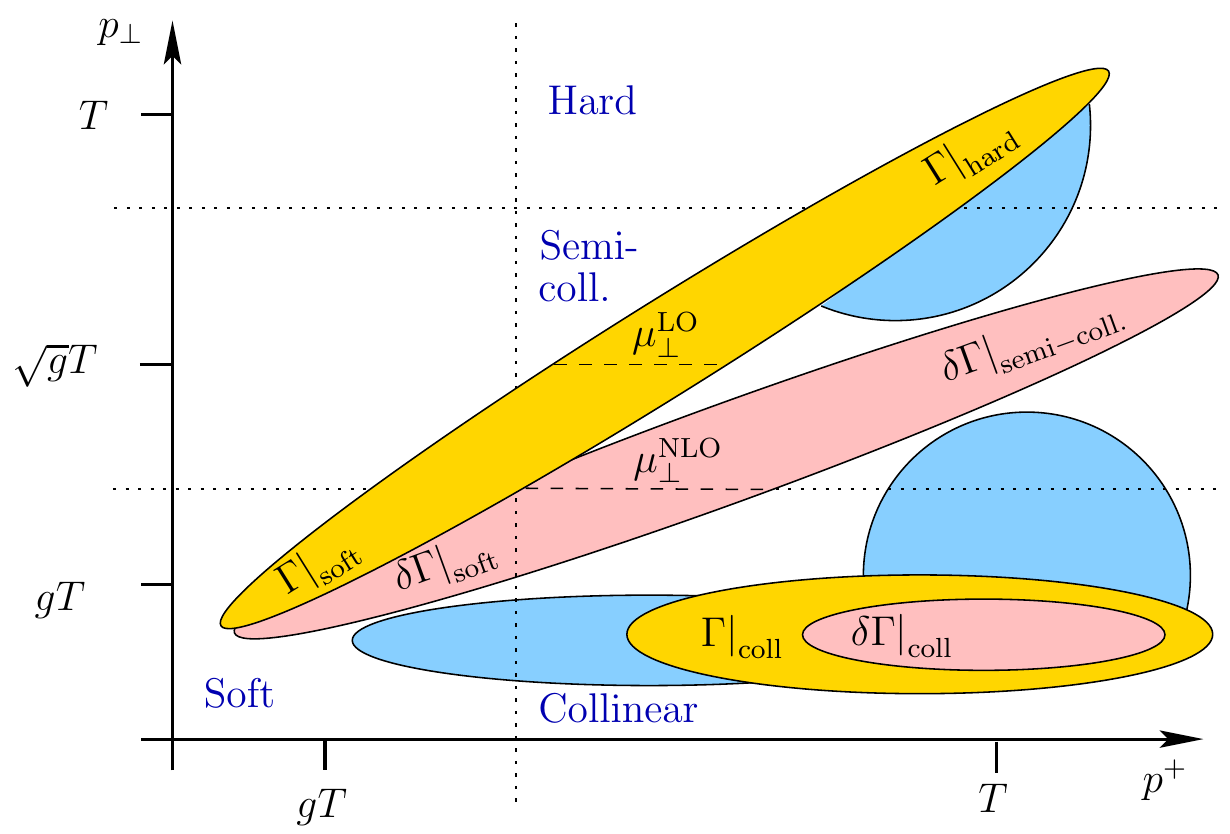}	
\end{center}
\caption{Regions contributing to the leading- (left) and next-to-leading order (right) calculations in the $(p^+,\pp)$ plane.}
\label{fig_map}	
\end{figure}

\begin{figure}[ht]
	\begin{center}
	\begin{minipage}{0.125\textwidth}
	\mbox{$\dgk_\mathrm{\hard}=$}
	\end{minipage}
	\begin{minipage}{0.65\textwidth}
	\includegraphics[width=10cm]{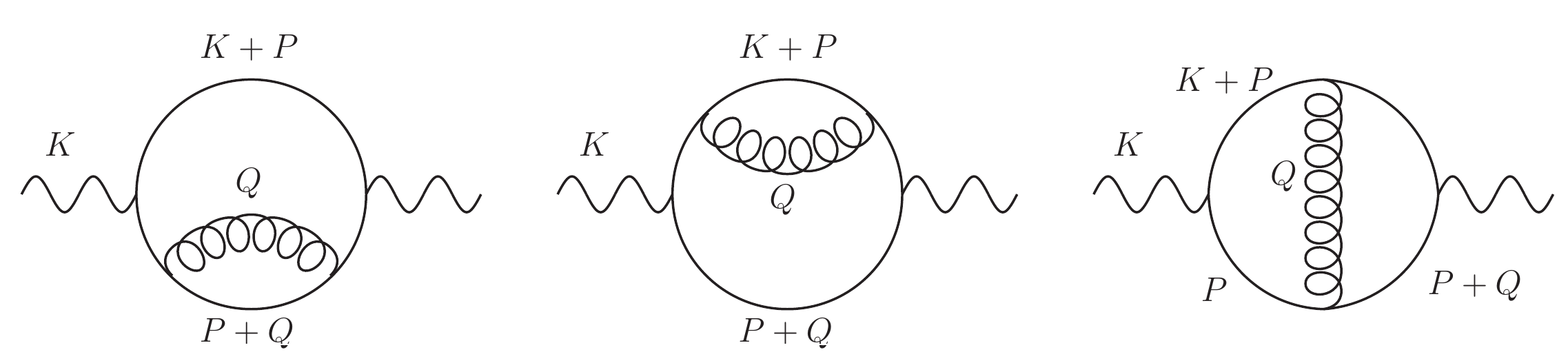}
	\end{minipage}
	\end{center}
	\vspace{-0.5cm}
	\caption{Two-loop diagrams necessary for the evaluation of the $\2to2$ region. 
	The wavy lines are photons, curly lines are gluons and plain lines are quarks.}
	\label{fig_diagrams}
\end{figure}

At the lowest order in pertubation theory Eq.~\eqref{defrate} vanishes, as the two bare Wightman
propagators in the simple quark loop it generates cannot be simultaneously put on shell. In
other words, on-shell quarks cannot radiate a physical photon. It is then necessary to
kick at least one of the quarks off-shell. This implies that the leading order is 
$\OO(\alpha g^2)$. Furthermore, the different kinematical regions contributing to the rate
are most naturally classified based on the scaling and virtuality. If $K$ is the photon momentum,
let us call $P$ and $K-P$ the fermion momenta at a current insertion in Eq.~\eqref{defrate}. Then,
as Fig.~\ref{fig_moms} schematically shows,
there are three relevant ways at LO to satisfy momentum conservation. In the \emph{hard} region one of the two is
far off-shell, $P^2\sim T^2$. In the \emph{soft} region $K-P$ carries basically the entire photon momentum and $P^2\sim g^2T^2$
is soft (all components of order $gT$).
 Finally, in the \emph{collinear} region, the three vectors are almost collinear and $P^2\sim g^2T^2$. We find
it convenient to choose $\k$ to point along $z$ and define light-cone coordinates $p^+\equiv (p^0+p^z)/2$, $p^-\equiv p^0-p^z$. 
The three scalings are then conveniently summarized in a $(p^+,\pp)$ plane, as shown on the left pane of Fig.~\ref{fig_map}.
As the figure suggests, the hard and soft regions are logarithmically sensitive to each other and a regulator ($\mu_\perp^\LO$)
is necessary in intermediate stages of the calculation.
In more detail, the hard region is dealt with by evaluating the simple two-loop diagrams in Fig.~\ref{fig_diagrams}, which, when cut, give
rise to the simple $qg\to q\gamma$ and $q\bar q\to g\gamma$ processes, folded over the thermal distributions of initial
and final states. The well known logarithmic IR divergences of these processes are cured by Hard Thermal Loop (HTL) resummation \cite{Braaten:1989mz}
in the soft sector, giving rise to a finite result for their sum \cite{Kapusta:1991qp,Baier:1991em}, which corresponds 
to the elongated diagonal shape in the left pane of Fig.~\ref{fig_map}.

\begin{figure}[ht]
	\begin{center}
	\begin{minipage}{0.08\textwidth}
		$\dgk_\mathrm{coll}$=
		\end{minipage}
		\begin{minipage}{0.75\textwidth}
		\includegraphics[width=13.2cm]{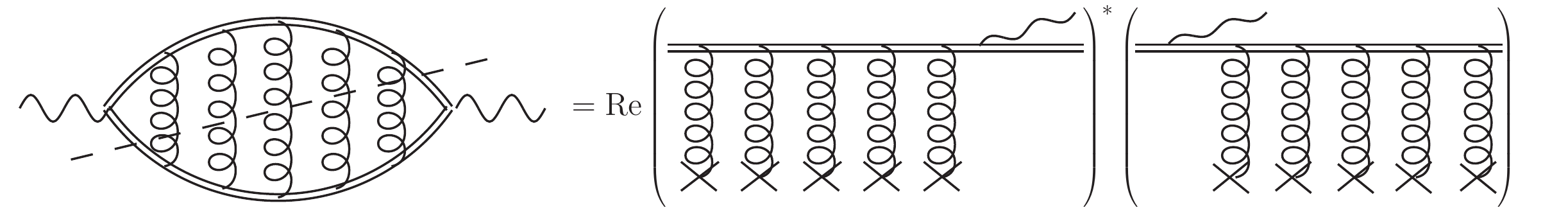}
		\end{minipage}
	\end{center}
	\caption{The  ladder diagrams that need to be resummed to account 
	for the LPM effect in the collinear region. The cut shown here corresponds to 
	the interference term on the right-hand side. The rungs on the l.h.s. are HTL 
	gluons in the Landau cut. On the r.h.s., the crosses at the 
	lower end of the gluons represent the hard scattering centers, either 
	gluons or fermions.}
	\label{fig_lpm}
\end{figure}

Collinear processes, as first pointed out in \cite{Aurenche:1998nw,Aurenche:1999tq,Aurenche:2000gf}, also contribute at leading order. The analysis and calculation
of \cite{Arnold:2001ba,Arnold:2001ms} showed how a soft, small-momentum transfer scattering with the 
medium constituents can broaden the transverse momentum of the emitting quarks just enough for  photon formation to be possible.
The long, $\OO(1/g^2T)$ formation time allows an arbitrary number of such scattering to contribute at leading order
and interfere, in what is called the \emph{Landau-Pomeranchuk-Migdal} (LPM) effect. Such an example is shown in Fig.~\ref{fig_lpm}.

The next-to-leading order ($\OO(\alpha g^3)$) comes from $\OO(g)$ corrections to this picture  that 
are of two kinds: \emph{loop corrections} and \emph{mistreated regions}.
The latter are $\OO(g)$ slices of the phase space of the LO calculation where intermediate states become soft without 
causing IR divergences. Integrating over these regions with unresummed matrix elements introduces an $\OO(g)$ error in the
LO calculation, which now needs to be addressed properly. This is done by identifying all such regions, extracting the limiting
behavior and subtracting it. We will not concentrate further on the details of this procedure.

Loop corrections arise instead from the addition of extra soft gluons, which are penalized by a factor of $g$ only due to
their Bose enhancement $T/p^0\sim T/(gT)$. The soft and 
collinear regions are both sensitive to these corrections, whereas they are by construction not possible in the hard region. Finally,
the slice of phase space corresponding to collinear scalings with larger virtuality ($P=(p^+,p^-,\pp)\sim(T,gT,\sqrt{g}T)$ rather than
$P\sim(T,g^2T,gT)$) needs to be treated with care, as it also contributes to NLO. We call it the \emph{semi-collinear} region. The NLO
contributions are summarized in the right pane of Fig.~\ref{fig_map}: the loop corrections are shown in salmon, whereas mistreated
regions are shown in blue.

\section{Summary of the calculation}
At leading order, the photon rate from the collinear sector reads \cite{Arnold:2001ba,Arnold:2001ms}
\begin{equation}
	\label{fpproblem}
 \dgk_{\coll}  =  \frac{ \alpha_{\rm EM}\nc }{2 k(2\pi)^3}
	  \sum_{s}  q_s^2
\int_{-\infty}^{+\infty} dp^+ \left[ \frac{(p^+)^2 + (p^+{+}k)^2}{(p^+)^2 (p^+{+}k)^2} \right]\:
   \nfd(k{+}p^+) [1-\nfd(p^+)]\:
\lim_{\x_\perp\to 0}{\rm Im}( 2 \nabla_{x_\perp} \cdot \f(\x_\perp)),
\end{equation}
which shows a nice factorization into a DGLAP-like splitting kernel (in square brackets), statistical (Fermi--Dirac)
functions for the emitters and finally their transverse size at the emission point. The latter
is determined through this Schr\"odinger-like inhomogeneous differential equation \cite{Arnold:2001ba,Arnold:2001ms,Aurenche:2002wq}
which resums the multiple soft scatterings, giving rise to their interference
\begin{equation}
\label{bspace}
-2i \nabla \delta^2(\x_\perp) = \frac{i k}{2p^+(k+p^+)} 
\Big( \mm - \nabla_{x_\perp}^2 \Big) \f(\x_\perp) + \cc(x_\perp) \f(\x_\perp) \,.
\end{equation}
This equation has two inputs, the thermal \emph{asymptotic mass} $m_\infty$ of the quarks and the \emph{scattering kernel}
$\cc(\xp)$. Since, intuitively, one has that if $p^+\gg\pp\gg p^-$, then in coordinate space $x^+\gg \xp\gg x^-$ and these
quantities can be defined through operators supported on the light cone along  $x^+$ or on the $x^-=0$ light front. For the former
one has \cite{CaronHuot:2008uw}
\begin{equation}
\label{defmm}
\mm=g^2\cf\left[\frac{1}{\da}\left\langle v_\mu F^{\mu\rho}
   \frac{-1}{(v\cdot D)^2} v_\nu F^\nu_{\,\rho}\right\rangle+\frac{1}{2\dr}\left\langle\overline\psi\frac{\slashed{v}}{v\cdot
    D}\psi\right\rangle\right]\stackrel{\LO}{=}g^2\cf\left[\frac{T^2}{6}+\frac{T^2}{12}\right]=\frac{g^2\cf T^2}{4},
\end{equation}
where $v=(1,{\bf v})$ is a null vector. For the latter instead one needs to consider a Wilson loop
in the $(x^+,\xp)$ plane \cite{CaronHuot:2008ni,Benzke:2012sz}, i.e.
	\begin{eqnarray}	
\label{defcc}
\cc(\xp) & = & \lim_{x^+\rightarrow \infty} (x^+)^{-1} \log\left[
	   \frac{1}{\nc}{\rm Tr}\: \Big\langle
	  \,U(0,\xp;x^+,\xp)  \,U(0,0;0,\xp) 
	   \, U(x^+,0;0,0)
	     \, U(x^+,\xp;x^+,0)\Big\rangle\right]_{x^-=\mathrm{const}}\\
\label{locc}
&\stackrel{\LO}{=}&g^2\cf T\int\frac{d^2\qp}{(2\pi)^2}\left(1-e^{i\x_\perp\cdot\bqp}\right)\left(\frac{1}{\qp^2}-\frac{1}{\qp^2+\md^2}\right)=\frac{g^2\cf T}{2\pi}
\big[K_0(\xp\md)+\gamma_\mathrm{E}+\ln(\xp\md/2)\big],
\end{eqnarray}
where $U$ is a straight, fundamental Wilson line and the leading-order expression comes from \cite{Aurenche:2002pd}.

Eq.~\eqref{fpproblem} is valid at NLO too; there, one needs to determine the $\OO(g)$ corrections to $\mm$ and $\cc(\xp)$.
A key observation, formulated in \cite{CaronHuot:2008ni} and based on the analytical properties, dictated by causality,
 of retarded and advanced
amplitudes, is that the thermal $n$-point function of $n$ fields on a
spacelike plane is given by the Euclidean Matsubara correlator with an imaginary, discrete component added to the spatial components
of the momenta. In the soft sector the lightlike limit can be reached smoothly and furthermore only the zero-mode is relevant at LO and NLO:
this implies that, for instance, the propagators become the standard, three-dimensional ones of dimensionally-reduced EQCD
\cite{Braaten:1995ju}. Indeed, the leading-order result in Eq.~\eqref{locc} is easily understood as the difference of the
massless transverse and massive longitudinal propagators at vanishing $q_z$. With this tremendous simplification\footnote{This 
simplification is also at the base of the recent lattice measurements of $\cc(\xp)$ and of the related jet-quenching parameter
$\hat{q}$ \cite{Laine:2013lia,Panero:2013pla,Marcotalk}} the NLO computations of $\cc(\xp)$ \cite{CaronHuot:2008ni} and $\mm$
\cite{CaronHuot:2008uw} become much simpler. Their results can be plugged in Eq.~\eqref{bspace} and treated as perturbations
to obtain the $\OO(g)$ correction to $\f(\x_\perp)$.

\begin{figure}[ht]
	\begin{center}
		\includegraphics[width=12cm]{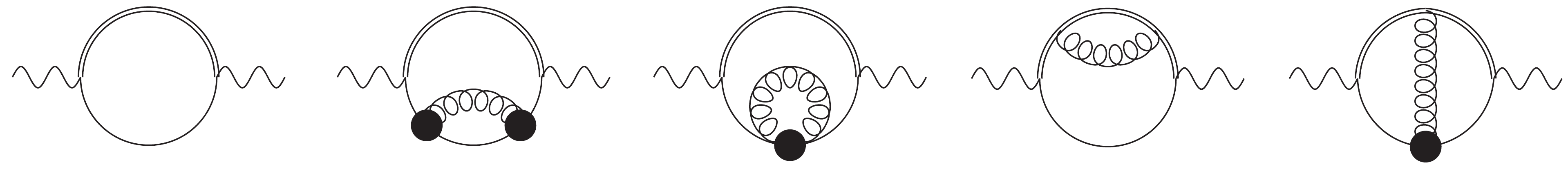}
	\end{center}
	\vspace{-0.5cm}
	\caption{Diagrams contributing to the soft rate at LO (first left) and NLO (others). The black blobs 
	are bare+HTL vertices, plain lines and gluons are soft, double line are hard quarks.}
	\label{fig_soft}
\end{figure}

\begin{figure}
\begin{center}
\includegraphics[width=8cm]{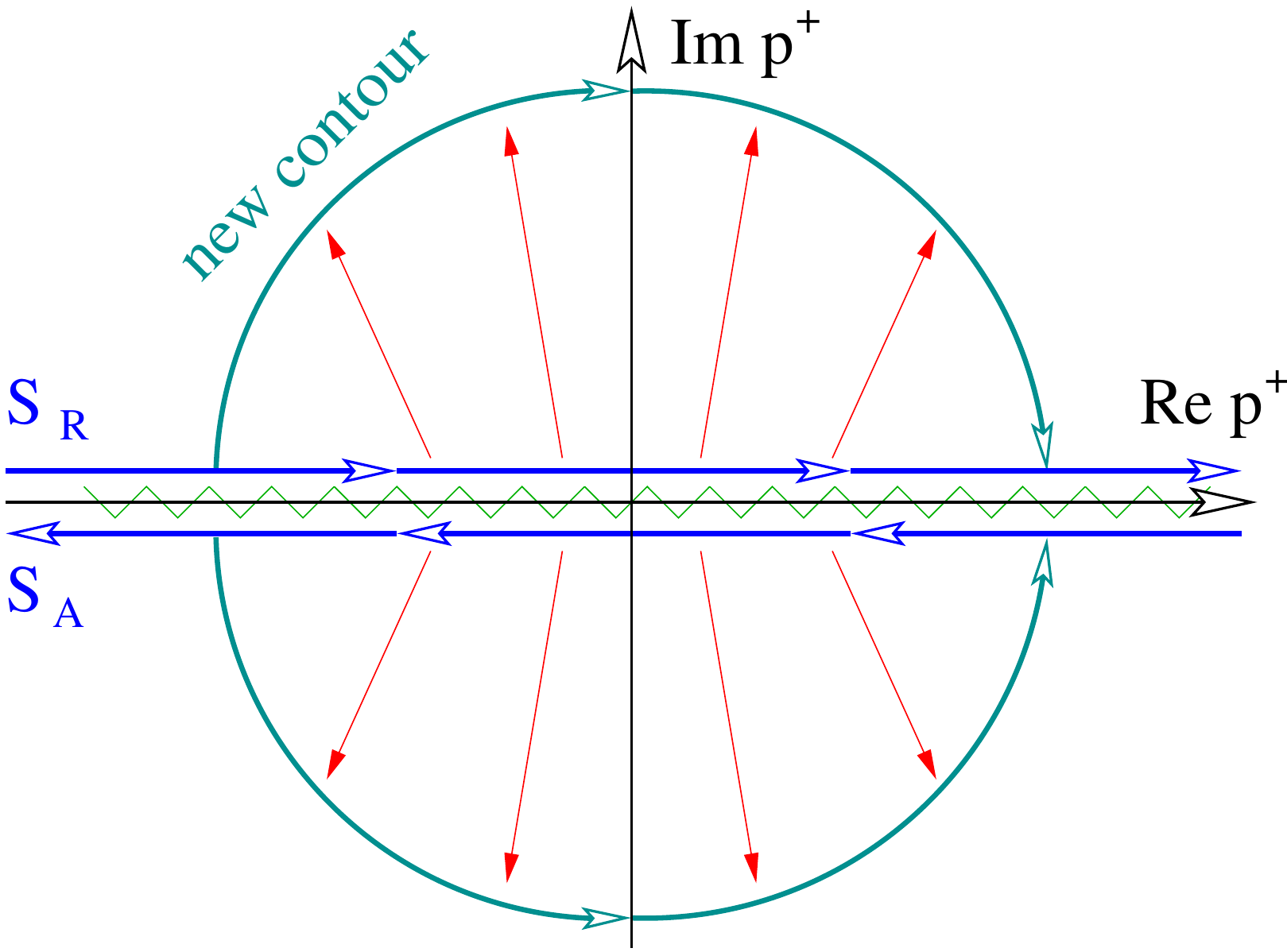}	
\put(-60,130){$i\Tr{\slashed{v}_kS_R}_{p^-=0}\to\displaystyle\frac{2\mm}{p^+(\pp^2+\mm)}$}
\end{center}
\vspace{-0.5cm}
\caption{Deformed contours for the evaluation of Eq.~\eqref{Wtrace}. Along those contours the integrand takes the simplified 
form shown along the upper arc.}
\label{fig_contour}	
\end{figure}

The observation at the base of this Euclideanization is that retarded correlators (e.g. a propagator) are analytical functions
in the upper half-plane in any timelike \emph{and lightlike} variable. This is of extreme importance for the soft
sector too. There one has at leading order
\begin{equation}
\dgk_{\soft}= -\frac{e^2}{2k(2\pi)^3}\sum_s q^2_s \dr \nfd(k) \int\frac{d^4P}{(2\pi)^4} 
   \Tr{ \slashed{v}_k (S_R(P)-S_A(P))}2\pi\delta(p^-),
	\label{Wtrace}
\end{equation}
which comes from the first diagram on the left in Fig.~\ref{fig_soft}.
At fixed $p^-$  the HTL-resummed retarded (advanced) soft quark propagator $S_R$ ($S_A$) is then guaranteed to be analytical
in the upper (lower) half-plane: we can then deform the contour away from the real axis to arcs at constant $\vert p^+\vert$, 
as shown in Fig.~\ref{fig_contour}.
$p^+$ is then effectively large, though complex, rendering the integrand considerably simpler.  The same can be done when evaluating
the four NLO diagrams in Fig.~\ref{fig_soft}, with their intricate HTL-resummed propagators and vertices. After subtracting the 
counterterm from the mistreated collinear limit we find that the NLO contribution arises only from the $\OO(g)$ $\delta \mm$ correction to
$\mm$, i.e.
\begin{equation}
i\,\Tr{\slashed{v}_kS_R(p^-{=}0)}\bigg\vert_\mathrm{arc}\stackrel{\LO}{=}\frac{2\mm}{p^+(\pp^2 + \mm)}\stackrel{\NLO}{\to}
\frac{2(\mm + \delta \mm)}{p^+(\pp^2 + \mm + \delta \mm)}
 =   \frac{2\mm}{p^+(\pp^2+\mm)} + 2\delta \mm \; \frac{\pp^2}{p^+(\pp^2+\mm)^2} \,,
\label{soft_guess}
\end{equation}
which also shows how both LO and NLO contributions are UV log-divergent. The former is cancelled by the hard region,
whereas the latter is compensated by the semi-collinear region.

For this latter region, we find that the diagrams of Fig.~\ref{fig_diagrams} (where the gluon is soft) 
are sufficient for its determination,
as LPM interference is suppressed. We find that the rate factorizes similarly to the collinear one, i.e.
\begin{equation}
	\label{semicoll}
	 \ddgkv_{\sc}  = 4\frac{ \alpha_{\rm EM} }{ k(2\pi)^3}
	  \sum_s \dr q_s^2
\int_{-\infty}^{+\infty}  dp^+ \left[ \frac{(p^+)^2 + (p^++k)^2}{k^2} \right]
 \nfd(k+p^+) [1-\nfd(p^+)]
\int \frac{d^2\pp}{(2\pi)^2}
  \,\frac{\hat{q}(\delta E)}{\pp^4},
  \end{equation}
where $\delta E=k\pp^2/(p^+(p^+{+}k))$ and $\hat{q}(\delta E)$ is a modified version of 
 $\hat{q}$ that keeps track of the evolving $p^-$ component of the momentum as well. It can 
 be defined through this light-cone operator
\begin{equation}
 \hat{q}(\delta E) =g^2\cf \int_{-\infty}^{+\infty} dx^+\,e^{ix^+\delta E} \,
    \frac{1}{\da} \langle v^\mu_{k} {F_\mu}^{\nu}(x^+,0,0_\perp)
       U_{\sss A}(x^+,0,0_\perp;0,0,0_\perp) 
                 v^\rho_{k}F_{\rho\nu}(0)\rangle,
\label{qhatde}
\end{equation}
where $v_k=K/k$. This operator can also be evaluated using Euclidean techniques and 
reverts to standard $\hat{q}$ for $\delta E\to0$.

\section{Results and conclusions}
\begin{figure}
	\includegraphics[width=0.495\textwidth]{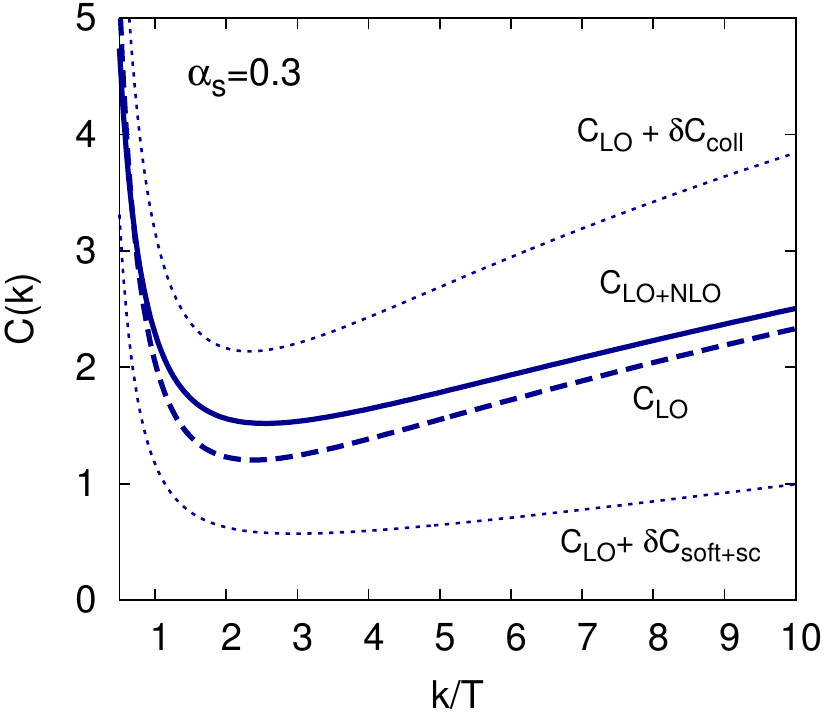}
	\includegraphics[width=0.495\textwidth]{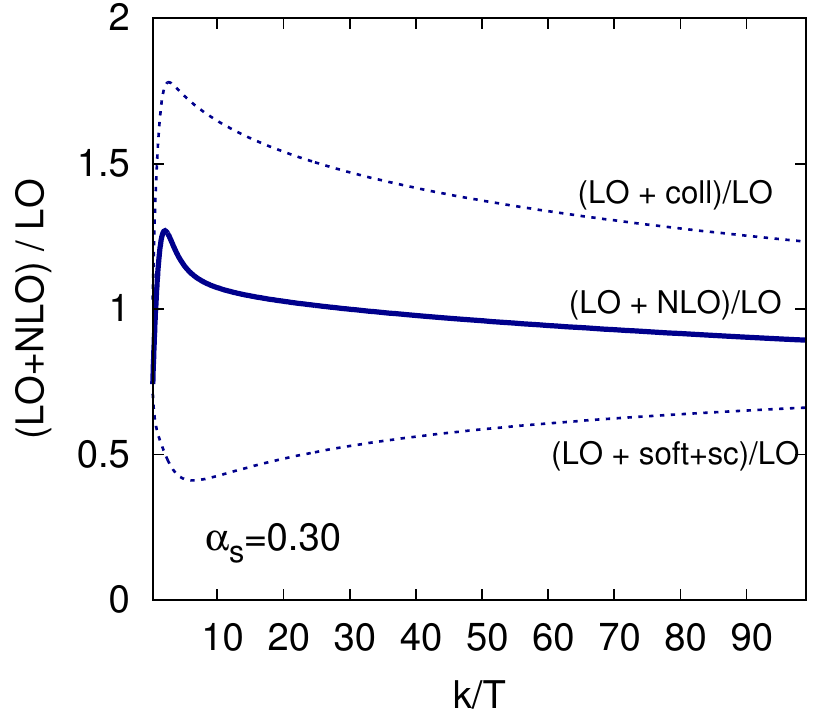}
   \caption{Left: the function, $C(k/T)\equiv d\Gamma/(d^3k)(4 \alpha_{\rm EM} \nfd(k) g^2 T^2/(3 k))^{-1} $, parametrizing the photon emission rate 
      for $\nc=\nf=3$ and $\als=0.3$.
       The full next-to-leading order function
         ($C_{\LO+\NLO}$) is a sum of the leading-order result ($C_{\LO}$), a 
         collinear correction ($\delta C_{\rm coll}$),  and a
         soft+semi-collinear correction ($\delta C_{\rm soft+sc}$).  
         The dashed curve  labeled
         $C_{\LO}+\delta C_{\rm coll}$ 
         shows the result when  only the collinear correction
         is included, with the analogous notation  for the  $C_{\LO} + \delta C_{\rm soft+sc}$ curve.  
		 The difference between the dashed curves provides a 
         uncertainty estimate  for the NLO calculation.
      Right: the ratio between the LO+NLO rate over the LO rate for the same parameters. }
	\label{fig_plot}
\end{figure}

In Fig.~\ref{fig_plot} we plot the rate (in units of the leading-log coefficient) and the ratio of LO+NLO correction over LO rate 
for $\nc=\nf=3$ and $\als=0.3$. Both plots show how in the phenomenologically relevant region the NLO correction represents an 
$\OO(20\%)$ increase which comes about from a cancellation between a much larger positive correction from the collinear sector
and a negative one from the semi-collinear and soft regions.
This cancellation appears largely accidental to us, confirmed by the fact that at larger momenta the negative contribution
overcomes the positive one. Hence, we believe that the band formed by these two curves can be taken as an uncertainty estimate of
the calculation. We refer to \cite{Ghiglieri:2013gia} for plots for smaller values of $\als$, which show a similar trend, as well
as for accurate fits of the results.

In conclusion, we have shown how the thermal photon rate can be computed at NLO in $g$ and how causality and analyticity play
a big role: in many cases the  computationally difficult physics of soft gluons and quarks can be factored into operators
supported on light fronts, which can be of two kinds. Either they are the correlators of the much simpler 3D theory, as for
$\cc(\xp)$, $\delta \mm$ and $\hat {q}(\delta E)$, or they can be mapped on arcs in the complex plane away from the real axis,
as for the soft quark contribution. An extension of these techniques to jet propagation and energy loss is underway\cite{jets};
 there we find 
that the longitudinal momentum diffusion coefficient can also be mapped on these arcs. 

\emph{Acknowledgements}
I thank Juhee Hong, Aleksi Kurkela, Egang Lu, Guy Moore and Derek Teaney for collaboration on this project.
 This work was supported by the Institute for Particle Physics (Canada) and the Natural
Sciences and Engineering Research Council (NSERC) of Canada.

\bibliographystyle{elsarticle-num}
\bibliography{photonbib.bib}







\end{document}